\documentstyle[eqsecnum,preprint,aps,prd,epsfig,amsfonts,amssymb]{revtex}
\tighten
\draft
\begin{document}
\preprint{OUTP-99-53-P:gr-qc/0004022}
\date{\today}

\title{THE RENORMALIZED STRESS TENSOR \\
IN KERR SPACE-TIME: GENERAL RESULTS }

\author{Adrian C. Ottewill} 
\address{Department of Mathematical Physics, University College Dublin,
\\ Belfield, Dublin 4, Ireland\\
email: ottewill@relativity.ucd.ie}

\author{Elizabeth Winstanley}
\address{Department of Theoretical Physics, University of Oxford,\\ 
1 Keble Road, Oxford OX1 3NP, UK\\
email: elizabeth.winstanley@oriel.ox.ac.uk}

\maketitle
\begin{abstract}
We derive constraints on the form of the renormalized stress tensor
for states on Kerr space-time based on general physical principles:
symmetry, the conservation equations, the trace anomaly and regularity
on (sections of) the event horizon.  This is then applied to the
physical vacua of interest.  
We introduce the concept of past and future Boulware
vacua and discuss the non-existence of a state empty at 
both  $\mathfrak{I}^-$ and $\mathfrak{I}^+$.
By calculating the stress tensor for the
Unruh vacuum at the event horizon and at infinity, we are able to
check our earlier conditions.  We also discuss the difficulties of
defining a state equivalent to the Hartle-Hawking vacuum and comment
on the properties of two candidates for this state.
\end{abstract}
 
\section{Introduction}
\label{sec:intro}
One of the central quantities of physical interest in a study of quantum field
theory in curved space-time is the renormalized expectation value of
the stress-energy tensor (RSET), since it is this quantity which
couples, via the semi-classical Einstein equations, to the background
geometry and thus yields the one-loop correction dynamics of the
geometry.  This paper is devoted to the properties of the RSET in the
states of greatest physical interest on Kerr space-times.  Any assault
on it by direct computation in black hole geometries is invariably a
long and complex process, requiring much algebraic dexterity and
ingenuity, and usually resorting to numerical analysis via computer.
The aim in this paper is to present what information can be gathered
from more physical principles and general considerations.  The most
important of these are the symmetries of the space-time and states
together with the conservation equations.  In addition, various
restrictions on the form of the RSET follow from its behaviour at the
event horizon and far from the black hole.  In a subsequent paper we
shall present numerical results for the RSET in the states appropriate
to a Kerr black hole with and without a bounding `box'.

The contents of this paper are as follows.  In section \ref{sec:wave}
we briefly review the solution of the wave equation in Kerr
space-time, concentrating for simplicity on the case of a conformally
coupled, massless scalar field.  We also introduce the standard
definitions of the Boulware and Unruh vacua, and discuss the
subtleties of defining the Hartle-Hawking vacuum in Kerr.  In the
absence of a true Hartle-Hawking state, we define two possible
candidates.  Next, in section \ref{sec:constraints} we investigate how
much information can be gathered about the stress tensor using the
conservation equations, symmetries of the geometry, and regularity
conditions on sections of the event horizon.  This greatly reduces the
number of unknown functions in the stress tensor.  The analysis of
this section is applicable to any quantum field, and any of the
physical vacua.  In section \ref{sec:asymptotics} we consider the
properties of the physical vacua in the asymptotic regions, at the
event horizon and at infinity, again concentrating on the massless
scalar field.  We calculate the differences in expectation values of
the stress tensor in the Unruh vacuum and other states, which can be
calculated without renormalization.  These calculations are in exact
agreement with our earlier analysis.  We also discuss the properties
of the candidate Hartle-Hawking states, in particular their symmetry
and regularity on the event horizon.

We follow the space-time conventions of Misner, Thorne and Wheeler
\cite{MTW} and work in geometric units throughout.

\section{The wave equation in Kerr space-time}
\label{sec:wave}
  
The Kerr line element in Boyer-Lindquist co-ordinates has the form
\begin{equation}
ds^2=-{\Delta \over \rho^2}(dt-a\sin^2\theta \, d\phi )^2
   + {\sin^2 \theta \over \rho^2} 
           \bigl((r^2+a^2)\, d\phi -a\, dt\bigr)^2
   +{\rho^2 \over \Delta}\, dr^2+\rho^2 \, d\theta ^2
\label{eq:metric}
\end{equation}
where
$\rho^2 =r^2+a^2\cos ^2 \theta$ 
and $\Delta =r^2-2Mr+a^2$.
Here $M$ is the mass of the black hole and $a$ its angular momentum 
per unit mass as viewed from infinity. 

The metric possesses two coordinate singularities at the roots of the
equation  $\Delta =0$, which we label 
$r=r_+=M+(M^2-a^2)^{1/2}$, defining the outer event horizon and
$r=r_-=M-(M^2-a^2)^{1/2}$, defining the inner Cauchy
horizon. In addition, there is a curvature singularity on the ring
defined by the equation  $\rho^2 =0$ (corresponding to
 $r=0$ and $\theta =\pi /2$).   

The space-time is stationary and axisymmetric, possessing two Killing
vectors, $\zeta =\partial /\partial t$ and $\eta =\partial /\partial
\phi $.  The former is timelike at infinity but becomes null when
$r=r_{s}=M+{\sqrt {M^2-a^2\cos ^2\theta }}$.  This surface is known as
the stationary limit surface and between it and the event horizon is a
region called the ergosphere.  Within the ergosphere, $\zeta $ is
spacelike and it is impossible for observers to remain at rest with
respect to infinity.  The stationary limit surface is timelike except
on the axis of symmetry $\theta =0$, where it joins the event horizon
and becomes null.  The Killing vector $\zeta + \Omega _{+}\eta $,
where $\Omega _{+}=a/(r_{+}^{2}+a^{2})=a/2Mr_{+}$ is the angular
velocity of the event horizon, generates the Killing horizon at
$r=r_{+}$.  This Killing vector is null on the event horizon, and
timelike outside it up to the velocity of light surface, at which
point it becomes null again.  The velocity of light surface is the
surface at which an observer with angular velocity $\Omega _{+}$ must
move with the speed of light.  It is not the same as the stationary
limit surface.  In addition, the space-time possesses a Killing-Yano
tensor which we shall discuss later.

Consider a conformally coupled massless scalar field satisfying the
equation $\partial _{\mu }(g^{1\over 2}g^{\mu\nu}\partial_{\nu })
\Phi=0$ (the scalar curvature $R$ being zero in Kerr space-time).
This equation is separable in the Kerr metric \cite{carter} and the
basis functions may be taken to be
\begin{equation}
 u_{\omega lm}(x)={N_{\omega lm}\over{(r^2+a^2)^{1\over 2}}}e^{-i\omega
t+im\phi} S_{\omega lm}(\cos \theta) R_{\omega lm}(r)
\label{eq:modes}
\end{equation}
 where $N_{\omega lm}$ is a normalization constant, $l$ and $m$ are
integers with $|m| \leq l$. $N_{\omega lm}$ is determined so that our
mode functions are orthonormal with respect to the standard 
inner product 
\begin{equation}
\langle u_1,u_2 \rangle ={1\over 2}i\int_{\Sigma }{\sqrt {-g}}\, 
(u_{2,\mu }^* u_1-u_2^*u_{1,\mu })\, d\Sigma ^{\mu }
\label{eq:inner_product}
\end{equation}
where $\Sigma $ is any Cauchy hypersurface. 

$S_{\omega lm}(\xi)$ is a
spheroidal harmonic satisfying the eigenvalue equation
\begin{equation}
{\left[ {d\over {d\xi }}(1-\xi ^2){d \over {d\xi }}-{m^2 \over
{1-\xi ^2}} +2ma\omega - (a\omega )^2(1-\xi ^2) +\lambda
_{lm}(a\omega ) \right] }S_{\omega lm} (\xi )=0 
\end{equation}
subject to regularity at $\xi = \pm 1$.
The eigenvalue $\lambda_{lm} (a\omega )$ depends on the integers
$l$ and $m$ and has the known value $\lambda_{lm}(0)=l(l+1)$, with
$S_{0 lm} (\xi )$ simply an associated Legendre function.
We may normalize the spheroidal harmonics so that
\begin{equation}
\int^1_{-1}S_{\omega lm}(\xi )S_{\omega l'm}(\xi )\, d\xi
=\delta_{ll'} .
\end{equation}

The radial equation may be written in the form
of a 1-dimensional time-independent Schr\"odinger equation
\begin{equation}
{\left[ {d^2\over {dr_{*}^{2}}}-V_{\omega lm}(r)\right] }R_{\omega
  lm}(r)=0 
\label{eq:2.2}
\end{equation}
where 
\begin{equation}
V_{\omega lm}(r)  = 
-\left( \omega -{ma\over {r^2+a^2}}\right)^2+
  \lambda_{lm}(a\omega){\Delta \over
  {(r^2+a^2)^2}}+{2(Mr-a^2)\Delta \over
  {(r^2+a^2)^3}}
+{3a^2\Delta ^2 \over {(r^2+a^2)^4}}\; ,
\label{eq:potential}
\end{equation}
and the `tortoise' co-ordinate $r_{*}$ is defined as
\begin{equation}
r_*=\int {r^2+a^2\over {\Delta }} \> dr
= r+ {1 \over 2 \kappa_+} \log |r-r_+|
+ {1 \over 2\kappa_-} \log |r-r_-| , 
\label{eq:tortoise}
\end{equation}
with
\begin{equation}
  \kappa_\pm =  {r_\pm - r_\mp \over 2 (r_\pm^2 + a^2)}  ,
\end{equation}
being the surface gravity on the inner and outer horizons.

In the asymptotic regions $r\to r_+$ 
($r_*\to -\infty $) and $r\to \infty$
($r_*\to \infty$) the potential (\ref{eq:potential}) reduces to
\begin{equation}
V_{\omega lm}(r) \sim
\cases{
 -(\omega -m\Omega_+)^2  &{\mbox{as   $r_*\to -\infty$}} \cr 
    -\omega^2  &{\mbox   {as $r_*\to \infty .$}}}
\end{equation}
We may thus choose as a basis of solutions to Eq.\ (\ref{eq:2.2}), two
classes of solutions with the asymptotic forms
\begin{eqnarray}
R^-_{\omega lm}(r) &\sim& \cases {
e^{i{\tilde {\omega }}r_*}+A^-_{\omega lm}e^{-i{\tilde {\omega }}r_*} &    
$ r_*\to -\infty $\cr
     B^-_{\omega lm}e^{i\omega r_*} & $ r_*\to \infty $\cr } 
\nonumber   \\
R^+_{\omega lm}(r) &\sim& \cases {B^+_{\omega lm}
e^{-i{\tilde {\omega }}r_*}  &
      $  r_*\to -\infty $\cr
e^{-i\omega r_*} + A^+_{\omega lm} e^{i\omega r_*}  & $    
r_*\to \infty$ \cr }
\label{eq:2.3}
\end{eqnarray}
where ${\tilde {\omega }}=\omega -m\Omega_+$. 
In the language of the Schr\"odinger equation analogy
it is natural to speak of $A$ and $B$ as the `reflection' and
`transmission' coefficients, respectively.

The eigenvalues $\lambda_{lm}$ are real and hence if $R$ is a 
solution of Eq.\ (\ref{eq:2.2}) then so too is $R^*$.  Using 
this and the constancy of the Wronskian for solutions to
Eq.\ (\ref{eq:2.2}) for 
various combinations of the radial
wavefunctions, it can be shown that  following relations hold 
\cite{dewitt}:
\begin{mathletters}
\begin{eqnarray}
1-|A^+_{\omega lm}|^2&=&{\omega -m\Omega_+\over {\omega }} 
|B^+_{\omega lm}|^2\\
1-|A^-_{\omega lm}|^2&=&{\omega \over {\omega -m\Omega_+}} 
|B^-_{\omega lm}|^2\\
\omega {B^-_{\omega lm}}^* A^+_{\omega lm}&=&
    -(\omega -m\Omega_+) B^+_{\omega lm}{A^-_{\omega lm}}^*\\
\omega B^-_{\omega lm}&=& (\omega -m\Omega_+) B^+_{\omega lm} .
\label{eq:2.4}
\end{eqnarray}
\end{mathletters}
The first two of these relations show that for $\omega >0$, $\omega
-m\Omega_+ = \tilde \omega <0$, both $|A^-|^2$ and $|A^+|^2$ are
greater than 1.  

\section{Quantum field theory in Kerr space-time}
\label{sec:qft}
\subsection{The mode functions}

We start by considering two natural complete, orthonormal sets of
solutions to the Klein-Gordon equation. It is then straightforward to
construct states with particular properties along a given Cauchy
surface, for example $\mathfrak{I}^- \cup \mathfrak{H}^-$.  Later, we
shall address the much more difficult question of constructing states
characterized on surfaces which do not form a Cauchy surface, for
example $\mathfrak{H}^- \cup \mathfrak{H}^+$.

With the understanding that $\omega >0$, we take as the `past' 
basis the following \cite{matacz}: 
\begin{mathletters}
\begin{eqnarray}
u_{\omega lm}^{in}&=&
\displaystyle{{1\over \sqrt{8\pi^2 \omega (r^2+a^2)}}}
e^{-i\omega t}e^{im\phi }S_{\omega lm}(\cos \theta ) R^+_{\omega
  lm}(r) \qquad \tilde\omega > - m \Omega_+
\label{eq:basis_in}  \\
u_{\omega lm}^{up}&=&
\displaystyle{{1 \over \sqrt{8\pi^2 {\tilde {\omega }}(r^2+a^2)}}}
e^{-i\omega t}e^{im\phi}S_{\omega lm}(\cos \theta )R^-_{\omega lm}(r)
\qquad \tilde\omega>0
\label{eq:basis_up}\\
u_{-\omega l-m}^{up}&=&\displaystyle{{1\over
 \sqrt{8\pi^2 (-{\tilde {\omega }})(r^2+a^2)}}}
e^{i\omega t}e^{-im\phi }S_{\omega lm}(\cos \theta) R^-_{-\omega l-m}(r) 
\qquad 0 > \tilde \omega > - m \Omega_+ 
\label{eq:basis_upS} 
\end{eqnarray} 
\label{eq:basis}
\end{mathletters}
where we have used the property
$S_{-\omega l-m}(\cos\theta)=S_{\omega lm}(\cos \theta)$.  
These  modes are orthonormal in the sense that
\begin{mathletters}
\begin{eqnarray}
(u_{\omega lm}^{in},u_{\omega' l'm'}^{in}) &=& \delta(\omega
-\omega')\delta_{ll'}\delta_{mm'} \qquad{\tilde {\omega}} > - m
\Omega_+\quad  [\omega>0]\\
(u_{\omega lm}^{up},u_{\omega' l'm'}^{up}) &=& \delta(\omega
-\omega')\delta_{ll'}\delta_{mm'} \qquad 
{\tilde {\omega }}>0\quad[\omega>m\Omega_+]\\
(u_{-\omega l-m}^{up},u_{-\omega' l'-m'}^{up}) &=& 
\delta(\omega -\omega')\delta_{ll'}\delta_{mm'} 
\qquad 0 > \tilde \omega >
- m \Omega_+\quad [m\Omega_+>\omega>0]
\end{eqnarray}
\end{mathletters}
with all other inner products vanishing.
Our conventions here adhere to those of the `distant observer
viewpoint' of Frolov and Thorne~\cite{frolov-thorne} which we will 
follow consistently throughout this series of papers.

From Eq.~(\ref{eq:2.3}),
\begin{mathletters}
\begin{eqnarray}
u^{in}_{\omega lm}&\sim& 
{S_{\omega lm}(\cos \theta )\over {\sqrt {8\pi ^2
\omega (r^2+a^2)}}}\times 
\cases{ 
0 & \mbox{at $\mathfrak{H}^-$} \cr 
\exp (-i\omega v + im\phi ) & \mbox{at $\mathfrak{I}^-$} \cr 
B^+_{\omega lm}\exp (-i{\tilde {\omega }} v+im\phi_+) & 
\mbox{at $\mathfrak{H}^+$} \cr
A^+_{\omega lm}\exp (-i\omega u+im\phi ) & 
\mbox{at $\mathfrak{I}^+$}
} 
\begin{array}{l}
{\tilde {\omega}} > - m \Omega_+
\\ \left[ \omega>0 \right]
\end{array}\\
u^{up}_{\omega lm}&\sim& 
{S_{\omega lm}(\cos \theta )\over {\sqrt {8\pi ^2
\tilde\omega (r^2+a^2)}}}\times 
\cases {
\exp (-i{\tilde {\omega }}u+im\phi_+) & at $\mathfrak{H}^-$ \cr
0 & at $\mathfrak{I}^-$ \cr
A_{\omega lm}^-\exp (-i{\tilde {\omega }}v+im\phi_+) & 
at $\mathfrak{H}^+$\cr
B_{\omega lm}^-\exp (-i\omega u+im\phi ) & at $\mathfrak{I}^+$ 
 } 
{
\begin{array}{l}
\tilde\omega > 0 \\ \left[ \omega>m\Omega_+ \right]
\end{array} }\\
u^{up}_{-\omega l-m}&\sim& 
{S_{\omega lm}(\cos \theta )\over {\sqrt {8\pi ^2
|{\tilde {\omega }}|(r^2+a^2)}}}\times 
\cases{
\exp (-i|{\tilde {\omega }}|u-im\phi_+) & at $\mathfrak{H}^-$ \cr
0 & at $\mathfrak{I}^-$ \cr
A_{\omega lm}^-\exp (-i|{\tilde {\omega }}|v-im\phi_+) & 
at $\mathfrak{H}^+ $\cr
B_{\omega lm}^-\exp (i\omega u-im\phi ) & at $\mathfrak{I}^+$ 
 }  
\begin{array}{l}
0 > \tilde\omega > -m\Omega_+
\\ \left[ m\Omega_+>\omega>0 \right]
\end{array}
\nonumber \\
\! \! \! \! & & 
\label{eq:null_asympupS}
\end{eqnarray}
\label{eq:null_asymp}
\end{mathletters}
where 
\begin{equation}
u=t-r_*, \quad v=t+r_*, \quad \phi_+ =\phi -\Omega_+ t.
\end{equation}
These modes are natural to the initial surfaces $\mathfrak{H}^-$ and
$\mathfrak{I}^-$ in the sense that $u^{in}$ describes unit flux coming
in from $\mathfrak{I}^-$ and zero flux coming up from
$\mathfrak{H}^-$, whereas $u^{up}$ describes unit flux coming up from
$\mathfrak{H}^-$ and zero incoming flux coming in from
$\mathfrak{I}^-$.  For modes with ${\tilde {\omega }}<0$ (but
$\omega>0$), $|A^-|^2>1$, so that they are reflected back to
$\mathfrak{H}^+$ with an amplitude greater than that they had
originally at $\mathfrak{H}^-$.  This is the classical phenomenon of
\emph{superradiance}.  Of course, as $\omega>0$ and $\Omega_+>0$ it is
only possible for ${\tilde {\omega }}=\omega - m\Omega_+$ to be
negative if $m>0$, that is for co-rotating waves.  Corresponding
comments apply to in modes with ${\tilde {\omega }}<0$: they are
reflected back to $\mathfrak{I}^+$ with an amplitude greater than that
they had originally at $\mathfrak{I}^-$.

Another aspect of superradiance is important to our discussion.  From
Eq. (\ref{eq:null_asympupS}), one sees that the up modes
(\ref{eq:basis_upS}) with $\tilde{\omega}<0$ have a negative energy
wave propagating to $\mathfrak{I}^+$ (conservation of energy). This is
a consequence of $\partial_t$ not being a globally time-like Killing
vector.  $\partial_t$ is space-like in the ergosphere, however the
combination $\partial_t + \Omega\partial_{\phi}$, where
$\Omega=-g_{t\phi}/g_{\phi\phi}$, is time-like down to the horizon
upon which it becomes null. Observers following integral curves of
this time-like vector field are locally non-rotating observers
(LNRO). A LNRO near the horizon would measure the frequency of the
superradiant up modes in (\ref{eq:basis_upS}) to be
$|\tilde\omega|=-\tilde{\omega}=-\omega+m\Omega_+$, in particular,
the LNRO would see positive frequency waves for all
modes. For $u^{in}_{\omega lm}$ all modes are positive frequency at
${\mathfrak {I}}^{+}$ and ${\mathfrak {I}}^{-}$. 
A LNRO near the horizon measures
$\tilde{\omega}$ for the frequency and thus sees negative frequency
modes in the superradiant regime.  An up mode having positive
frequency with respect to $u$ at $\mathfrak{H}^-$ will have negative
frequency with respect to $u$ at $\mathfrak{I}^+$ if ${\tilde {\omega
}}<0$ but $\omega>0$.

\subsection{The Physical Vacua}
\label{subsec:vacuums}

We now turn to the delicate issue of defining analogs of the standard
three vacuum states in Schwarzschild space-time (Boulware,
Hartle-Hawking and Unruh) in Kerr space-time.  (Our discussion here
concerns states on the full exterior region of Kerr, in later papers
we shall
also talk about the case when the black hole is contained within a
`box'.)  The construction of vacuum states in Kerr is a more subtle
problem than for Schwarzschild black holes, for the following reasons:
\begin{enumerate}
\item 
The existence of superradiant modes makes the definition of
positive frequency more complicated.  For example, in Schwarzschild,
an outgoing mode which has positive frequency with respect to the
retarded null co-ordinate $u$ at the past horizon ${\mathfrak
{H}}^{-}$ will also have positive frequency with respect to $u$ at
${\mathfrak {I}}^{+}$, so it does not matter if we define positive
frequency with respect to $u$ at ${\mathfrak {H}}^{-}$ or
at ${\mathfrak {I}}^{+}$.  This is no longer the case in Kerr:  a
superradiant mode can have positive frequency with respect to
$u$ at ${\mathfrak {I}}^{+}$ but negative frequency at ${\mathfrak
{H}}^{-}$.  This is why our definition of the basis of mode functions
(\ref{eq:basis}) had to be so carefully done.
\item
As a consequence of this, it is \emph{only} straightforward to define
states with particular properties along a given Cauchy surface, such
as ${\mathfrak {I}}^{-}\cup {\mathfrak {H}}^{-}$.  By contrast, it has
become conventional in Schwarzschild space-time to consider the
Boulware vacuum in terms of its properties on ${\mathfrak {I}}^{-}\cup
{\mathfrak {I}}^{+}$ and the Hartle-Hawking vacuum in terms of its
properties on ${\mathfrak {H}}^{-}\cup {\mathfrak {H}}^{+}$.
\end{enumerate}

To be explicit, we may expand the scalar field $\Phi (x)$ 
in terms of the mode functions we introduced above
\begin{eqnarray}
\Phi (x)&=&\sum_{l,m} \left(
\int_{0}^{\infty }d\omega \, 
(a^{in}_{\omega lm}
  u_{\omega lm}^{in}+a^{in\dag }_{\omega lm}u^{in*}_{\omega lm})
+ \int_{\omega_{min}}^{\infty }d\omega \, 
(a^{up}_{\omega lm}
  u_{\omega lm}^{up}+a^{up\dag}_{\omega lm}u^{up*}_{\omega lm})
 \right) \nonumber \\
&&+\sum_{l,m} 
\int_{0}^{\omega_{min}}d\omega \, 
(a^{up}_{-\omega l-m}
  u_{-\omega l-m}^{up}+a^{up\dag }_{-\omega l-m}
u^{up*}_{-\omega l-m})
\nonumber \\
&=&\sum_{l,m} \left(
\int_{0}^{\infty }d\omega \, 
(a^{in}_{\omega lm}
  u_{\omega lm}^{in}+a^{in\dag }_{\omega lm}u^{in*}_{\omega lm})
+ \int_0^{\infty }d\tilde \omega \, 
(a^{up}_{\omega lm}
  u_{\omega lm}^{up}+a^{up\dag}_{\omega lm}u^{up*}_{\omega lm})
\right)
\end{eqnarray}
where $\omega_{min}=\max \lbrace 0,m\Omega_+\rbrace $, so
$\omega_{min} = 0$ for counter-rotating waves ($m \leq 0$) 
and $\omega_{min} = m\Omega_+$ for co-rotating waves ($m > 0$). 
Given this expansion, the natural way  to quantize the field is
for  the coefficients to  become operators satisfying 
the commutation relations
\begin{mathletters}
\begin{eqnarray}
\left[{\hat a}^{in}_{\omega lm}, {\hat a}^{in\dag }_{\omega 'l'm'}\right] &=& 
\delta (\omega -\omega')\delta_{ll'}\delta_{mm'}
 \qquad{\tilde {\omega}} > - m \Omega_+\\
\left[{\hat a}^{up}_{\omega lm},  {\hat a}^{up\dag }_{\omega 'l'm'}\right] &=&
 \delta(\omega-\omega')\delta_{ll'}\delta_{mm'} 
\qquad \tilde\omega>0\\
\left[{\hat a}^{up}_{-\omega l-m},{\hat a}^{up\dag }_{-\omega 'l'-m'} \right] &=& 
\delta(\omega -\omega')\delta_{ll'}\delta_{mm'}
 \qquad 0 > \tilde \omega > - m \Omega_+
\end{eqnarray}
\end{mathletters}
with all other commutators vanishing.  From Eq.(\ref{eq:null_asymp}),
the operators ${\hat a}^{in\dag }$ and ${\hat a}^{up\dag}$ have
the natural interpretation that they will, respectively, create
particles incident from $\mathfrak{I}^-$ and $\mathfrak{H}^-$.  With
this in mind, we define a `past Boulware' vacuum state by
\begin{mathletters}
\begin{eqnarray}
{\hat a}_{\omega lm }^{in} | B^{-} \rangle 
&=& 0 \qquad \tilde\omega > - m\Omega_+\\
{\hat a}_{\omega lm}^{up} | B^{-} \rangle 
&=& 0 \qquad \tilde\omega>0 \\
{\hat a}_{-\omega l-m}^{up} | B^{-} \rangle 
&=& 0 \qquad 0 > \tilde\omega > - m
\Omega_+
\end{eqnarray}
\end{mathletters}
corresponding to an absence of particles from 
${\mathfrak{H}}^{-}$ and ${\mathfrak{I}}^{-}$.

This state does not precisely correspond to the idea of
a Boulware state in Schwarzschild as that state which is
most empty at infinity. The state $|B^{-}\rangle $ contains, 
at ${\mathfrak {I}}^{+}$, an outward flux of particles in the 
superradiant modes; this is the Unruh-Starobinskii effect 
\cite{unruh}.

One might suppose that a more appropriate definition for
the Boulware vacuum would be to define a state which is
empty at ${\mathfrak {I}}^{-}$ and ${\mathfrak {I}}^{+}$.
However, it is straightforward to see that such a state
cannot exist within conventional quantum field theory by introducing 
the mode functions natural for defining the `future Boulware' vacuum.
(We shall discuss later the non-conventional 
`$\eta$-formalism' construction
proposed by Frolov and Thorne\cite{frolov-thorne}.)

The mode functions relevant to the `future Boulware' vacuum are
those representing a unit (locally-positive frequency) flux out to  
$\mathfrak{I}^{+}$ and down $\mathfrak{H}^{+}$.
From the asymptotic forms for the radial functions Eq. (\ref{eq:2.3}),
it is clear that we should take as our `future' basis \cite{matacz}:
\begin{mathletters}
\begin{eqnarray}
u_{\omega lm}^{out}&=&
\displaystyle{{1\over \sqrt{8\pi^2 \omega (r^2+a^2)}}}
e^{-i\omega t}e^{im\phi }S_{\omega lm}(\cos \theta ) R^{+*}_{\omega
  lm}(r) \qquad \tilde\omega > - m \Omega_+ ,
\label{eq:basis_out}  \\
u_{\omega lm}^{down}&=&\displaystyle{{1 \over \sqrt{8\pi^2 
{\tilde {\omega }}(r^2+a^2)}}}
e^{-i\omega t}e^{im\phi}S_{\omega lm}(\cos \theta )
R^{-*}_{\omega lm}(r)
\qquad \tilde\omega>0 ,
\label{eq:basis_down}\\
u_{-\omega l-m}^{down}&=&\displaystyle{{1\over
 \sqrt{8\pi^2 |{\tilde {\omega }}|(r^2+a^2)}}}
e^{i\omega t}e^{-im\phi }S_{\omega lm}(\cos \theta) 
R^{-*}_{-\omega l-m}(r) 
\qquad 0 > \tilde \omega > - m \Omega_+ .
\label{eq:basis_downS} 
\end{eqnarray} 
\end{mathletters}
These  modes are orthonormal in the sense that
\begin{mathletters}
\begin{eqnarray}
(u_{\omega lm}^{out},u_{\omega' l'm'}^{out}) &=& \delta(\omega
-\omega')\delta_{ll'}\delta_{mm'} \qquad{\tilde {\omega}} > - m
\Omega_+\quad  [\omega>0]\\
(u_{\omega lm}^{down},u_{\omega' l'm'}^{down}) &=& \delta(\omega
-\omega')\delta_{ll'}\delta_{mm'} \qquad 
{\tilde {\omega }} >0\quad[\omega>m\Omega_+] \\
(u_{-\omega l-m}^{down},u_{-\omega' l'-m'}^{down}) &=& 
\delta(\omega -\omega')\delta_{ll'}\delta_{mm'} \qquad 
0 > \tilde \omega >
- m \Omega_+\quad [m\Omega_+>\omega>0]
\end{eqnarray}
\end{mathletters}
with all other inner products vanishing.
Their asymptotic properties are given by 
\begin{mathletters}
\begin{eqnarray}
u^{out}_{\omega lm}&\sim& 
{S_{\omega lm}(\cos \theta )\over {\sqrt {8\pi ^2
\omega (r^2+a^2)}}}\times 
\cases{ 
B^{+*}_{\omega lm}\exp (-i{\tilde\omega}u+im\phi_+) 
& \mbox{at $\mathfrak{H}^-$} \cr 
A^{+*}_{\omega lm}\exp (-i\omega v + im\phi ) 
& \mbox{at $\mathfrak{I}^-$} \cr 
0 & \mbox{at $\mathfrak{H}^+$} \cr
\exp (-i\omega u+im\phi ) & \mbox{at $\mathfrak{I}^+$}
 } 
\begin{array}{l}
{\tilde {\omega}} > - m \Omega_+
\\ \left[ \omega>0 \right]
\end{array}
\\
u^{down}_{\omega lm}&\sim& 
{S_{\omega lm}(\cos \theta )\over {\sqrt {8\pi ^2
\tilde\omega (r^2+a^2)}}}\times 
\cases {
A^{-*}_{\omega lm}\exp(-i{\tilde\omega}u+im\phi_+) 
& \mbox{at $\mathfrak{H}^-$} \cr
B^{-*}_{\omega lm}\exp (-i\omega v + im\phi ) 
& \mbox{at $\mathfrak{I}^-$} \cr
\exp(-i{\tilde\omega}v+im\phi_+) & \mbox{at $\mathfrak{H}^+$}\cr
0 &\mbox{at $\mathfrak{I}^+$} 
} 
\begin{array}{l}
\tilde\omega > 0 \\ \left[ \omega>m\Omega_+ \right]
\end{array} 
\\
u^{down}_{-\omega l-m}&\sim& 
{S_{\omega lm}(\cos \theta )\over {\sqrt {8\pi ^2
|{\tilde {\omega }}|(r^2+a^2)}}}\times 
\cases{
A^{-*}_{-\omega l-m}\exp (-i|{\tilde\omega}|u-im\phi_+) 
& \mbox{at $\mathfrak{H}^-$} \cr
B^{-*}_{-\omega l-m}\exp (i\omega v-im\phi) 
& \mbox{at $\mathfrak{I}^-$} \cr
\exp (-i|{\tilde\omega}|v-im\phi_+) & 
\mbox{at $\mathfrak{H}^+ $}\cr
0 &\mbox{ at $\mathfrak{I}^+$} 
 }  
\begin{array}{l}
0 > \tilde\omega > -m\Omega_+
\\ \left[ m\Omega_+>\omega>0 \right]
\end{array}
\nonumber \\ 
\! \! \! \! & &
\label{eq:null_asympdownS}
\end{eqnarray}
\label{eq:null_asymp+}
\end{mathletters}

We may expand the scalar field $\Phi (x)$ 
in terms of these mode functions we introduced above
\begin{eqnarray}
\Phi (x)&=&\sum_{l,m} \left(
 \int_{0}^{\infty }d\omega \, 
(a^{out}_{\omega lm}
  u_{\omega lm}^{out}+a^{out\dag }_{\omega lm}u^{out*}_{\omega lm})
+  \int_{\omega_{min}}^{\infty }d\omega \, 
(a^{down}_{\omega lm}
  u_{\omega lm}^{down}+a^{down\dag}_{\omega lm}u^{down*}_{\omega lm})
\right)  
\nonumber \\
&&+\sum_{l,m}
\int_{0}^{\omega_{min}}d\omega \, 
(a^{down}_{-\omega l-m}
  u_{-\omega l-m}^{down}+a^{down\dag}_{-\omega l-m}
u^{down*}_{-\omega l-m}) \nonumber \\
&=&\sum_{l,m}\left(
 \int_{0}^{\infty }d\omega \, 
(a^{out}_{\omega lm}
  u_{\omega lm}^{out}+a^{out\dag }_{\omega lm}u^{out*}_{\omega lm})
+
 \int_0^{\infty }d\tilde\omega \, 
(a^{down}_{\omega lm}
  u_{\omega lm}^{down}+a^{down\dag}_{\omega lm}u^{down*}_{\omega lm})
\right) 
\end{eqnarray}
where $\omega_{min}=\max \lbrace 0,m\Omega_+\rbrace $, as before.
Given this expansion, the natural way  to quantize the field is
for  the coefficients  become operators satisfying 
the commutation relations
\begin{mathletters}
\begin{eqnarray}
\left[{\hat a}^{out}_{\omega lm}, 
{\hat a}^{out\dag }_{\omega 'l'm'}\right] &=&
\delta (\omega -\omega')\delta_{ll'}\delta_{mm'}
 \qquad{\tilde {\omega}} > - m \Omega_+\\
\left[{\hat a}^{down}_{\omega lm},  
{\hat a}^{down\dag }_{\omega 'l'm'}\right] &=&
 \delta(\omega-\omega')\delta_{ll'}\delta_{mm'} 
\qquad \tilde\omega>0\\
\left[{\hat a}^{down}_{-\omega l-m},
{\hat a}^{down\dag }_{-\omega 'l'-m'} \right] &=& 
\delta(\omega -\omega')\delta_{ll'}\delta_{mm'}
 \qquad 0 > \tilde \omega > - m \Omega_+
\end{eqnarray}
\end{mathletters}
with all other commutators vanishing.  From Eq.(\ref{eq:null_asymp+}),
the operators ${\hat a}^{out\dag }$ and ${\hat a}^{down\dag}$ have
the natural interpretation that they will, respectively, create
particles incident from $\mathfrak{I}^+$ and $\mathfrak{H}^+$. Thus,
we define the `future Boulware' vacuum state by
\begin{mathletters}
\begin{eqnarray}
{\hat a}_{\omega lm }^{out} | B^{+} \rangle &=& 
0 \qquad \tilde\omega >  - m\Omega_+\\
{\hat a}_{\omega lm}^{down} | B^{+} \rangle &=& 
0 \qquad \tilde\omega>0 \\
{\hat a}_{-\omega l-m}^{down} | B^{+} \rangle &=& 
0 \qquad 0 > \tilde\omega > - m\Omega_+
\end{eqnarray}
\end{mathletters}
corresponding to an absence of particles from 
${\mathfrak{H}}^{+}$ and ${\mathfrak{I}}^{+}$.
In this language, the Unruh-Starobinskii effect is a statement
about the behaviour of
\begin{equation}
     \langle B^- | \hat T_{\mu\nu} | B^- \rangle - 
          \langle B^+ | \hat T_{\mu\nu} | B^+ \rangle
\end{equation}
as $r \to \infty$.

A vacuum state empty at $\mathfrak{I}^-$ and $\mathfrak{I}^+$ must be
constructed from modes $u^{in}_{\omega lm}$ and $u^{out}_{\omega lm}$
up to a trivial Bogoliubov transformation (i.e., one with all
$\beta$-coefficients vanishing).  However, $u^{in}_{\omega lm}$ and
$u^{out}_{\omega lm}$ are not orthogonal and the fact that they cannot
be made so by any trivial Bogoliubov transformation is seen most
easily by writing $u^{out}_{\omega lm}$ in terms of the basis given by
$u^{in}_{\omega lm}$ and $u^{up}_{\omega lm}$.  For non-superradiant
modes the transformation does correspond to a trivial Bogoliubov
transformation:
\begin{mathletters}
\begin{eqnarray}
  \label{eq:ns_bog}
  u^{out}_{\omega lm} &=& A^{+*}_{\omega lm}u^{in}_{\omega lm} +
 \sqrt{\tilde \omega \over 
 \omega} B^{+*}_{\omega lm} u^{up}_{\omega
  lm}, \qquad\tilde\omega > 0  ,\\
 u^{down}_{\omega lm} &=&  \sqrt{\omega \over  \tilde \omega} 
B^{-*}_{\omega lm} u^{in}_{\omega  lm} + 
A^{-*}_{\omega lm}u^{up}_{\omega lm} , \qquad [\omega>m\Omega_+] ,
\end{eqnarray}
\end{mathletters}
but for superradiant modes
\begin{mathletters}
\begin{eqnarray}
  \label{eq:s_bog}
  u^{out}_{\omega lm} &=& A^{+*}_{\omega lm}u^{in}_{\omega lm} -
\sqrt{-\tilde\omega \over \omega} 
B^{+*}_{\omega lm}u^{up*}_{-\omega l-m},
 \qquad 0 > \tilde\omega > -m\Omega_+ 
\left[ m\Omega_+>\omega>0 \right] ,\\
 u^{down}_{-\omega l-m} &=&  \sqrt{\omega \over -\tilde
 \omega} B^{-*}_{-\omega l-m} u^{in*}_{\omega
  lm} + A^{-*}_{-\omega l-m} u^{up}_{\omega lm} , \qquad
0 > \tilde\omega > -m\Omega_+ \left[ m\Omega_+>\omega>0 \right] .
\end{eqnarray}
\end{mathletters}
As no trivial Bogoliubov transformation can affect the total number
of `particles' produced, $\sum_{i,r} |\beta_{ir}|^2$, it is impossible
to define a vacuum state empty with respect to in modes at
$\mathfrak{I}^-$ and out modes at $\mathfrak{I}^+$.

The non-existence of a `true Boulware' state is intimately 
linked with the
non-existence a `true Hartle-Hawking' state (defined as being
a Hadamard state which respects the symmetries of the space-time and
is regular everywhere, in particular, on both future and past event
horizons) on Kerr space-time \cite{kay-wald}.  In the former case,
one wishes to define the state on  $\mathfrak{I}^- \cup
\mathfrak{I}^+$, in the latter on $\mathfrak{H}^- \cup
\mathfrak{H}^+$.  Indeed, one can make the analogy quite precise by,
in the language of Frolov and Thorne, switching from a `distant' to a 
`near horizon' viewpoint. 

The (past) Unruh state $|U^{-}\rangle $ is easily defined as that
state empty at ${\mathfrak {I}}^{-}$ but with the `up' modes
(natural modes on ${\mathfrak {H}}^{-}$) thermally populated.  
For a proof that this is equivalent to using modes which are 
positive frequency with respect to a future-increasing affine parameter
on ${\mathfrak {H}}^{-}$ see Ref.~\cite{frolov-thorne}.
As before, we use the notation $|U^{-}\rangle $ in order
to emphasize that this state is naturally defined by considerations on 
${\mathfrak {H}}^{-} \cup {\mathfrak {I}}^{-}$.
One can, of course also define a state $|U^{+}\rangle $ empty at 
${\mathfrak {I}}^{+}$ but with the `down' modes
(natural modes on ${\mathfrak {H}}^{+}$) thermally populated. 
Indeed, one can also make such a distinction in the Schwarzschild
case for the Unruh vacuum.  However, one rarely considers 
$|U^{+}\rangle$
as it is $|U^{-}\rangle$ that mimics the state arising at late times
from the collapse of a star to a black hole. 
For this reason we shall usually
drop the term `past' but we will retain the terminology $|U^{-}\rangle
$ to make clear that this state is naturally defined in terms of `in'
and `up' modes. In this language, the (Kruskal space-time model of
the) Hawking effect is a statement about the behaviour of
\begin{equation}
     \langle U^- | \hat T_{\mu\nu} | U^- \rangle - 
          \langle B^+ | \hat T_{\mu\nu} | B^+ \rangle
\end{equation}
as $r \to \infty$.

With these definitions, it is straightforward to write down mode
sum expressions for the two-point functions of the field in
the past and future Boulware and (past) Unruh vacuum states:
\begin{mathletters}
\begin{eqnarray}
G_{B^-}(x,x') &=& \langle B^-|\hat \Phi(x) 
\hat \Phi(x')|B^-\rangle \nonumber\\
&=& \sum_{l,m}\left(
\int_{0}^{\infty }d{\tilde {\omega }}\, 
u_{\omega lm }^{up}(x)u_{\omega lm}^{up*}(x')
+\int_{0}^{\infty }d\omega \, 
u_{\omega lm }^{in}(x)u_{\omega lm}^{in*}(x')\right)
\label{eq:b-_green}\\
G_{B^+}(x,x') &=& \langle B^+|\hat \Phi(x) \hat \Phi(x')|B^+\rangle 
\nonumber\\
&=& \sum_{l,m}\left(
\int_{0}^{\infty }d{\tilde {\omega }}\, 
u_{\omega lm }^{down}(x)u_{\omega lm}^{down*}(x')
+\int_{0}^{\infty }d\omega \, 
u_{\omega lm }^{out}(x)u_{\omega lm}^{out*}(x')\right)
\label{eq:b+_green}\\
G_{U^-}(x,x') &=& \langle U^-|\hat \Phi(x) \hat \Phi(x')|U^-\rangle 
\nonumber\\
&=& \sum_{l,m}\left(
\int_{0}^{\infty }d{\tilde {\omega }}\, {\hbox {coth}}
\left( {\pi {\tilde {\omega }}\over {\kappa }}\right) 
u_{\omega lm }^{up}(x)u_{\omega lm}^{up*}(x)
 +\int_{0}^{\infty }d\omega \, 
u_{\omega lm }^{in}(x)u_{\omega lm}^{in*}(x')\right) .
\label{eq:u-_green}
\end{eqnarray}
\end{mathletters}

The corresponding expressions for the  
unrenormalized expectation values of 
the stress tensor in the past and future Boulware and 
(past) Unruh vacuum states are:
\begin{mathletters}
\begin{eqnarray}
\langle B^-|{\hat {T}}_{\mu\nu}|B^-\rangle 
& = & \sum_{l,m}\left(
\int_{0}^{\infty }d{\tilde {\omega }}\, 
T_{\mu\nu}[u_{\omega lm }^{up},u_{\omega lm}^{up*}]
+\int_{0}^{\infty }d\omega \, 
T_{\mu\nu}[u_{\omega lm }^{in},u_{\omega lm}^{in*}]\right)
\label{eq:b-}\\
\langle B^+|{\hat {T}}_{\mu\nu}|B^+\rangle 
& = & \sum_{l,m}\left(
\int_{0}^{\infty }d{\tilde {\omega }}\, 
T_{\mu\nu}[u_{\omega lm }^{down},u_{\omega lm}^{down*}]
+\int_{0}^{\infty }d\omega \, 
T_{\mu\nu}[u_{\omega lm }^{out},u_{\omega lm}^{out*}]\right)
\label{eq:b+}\\
\langle U^-|{\hat {T}}_{\mu\nu}|U^-\rangle 
& = & \sum_{l,m}\left(
\int_{0}^{\infty }d{\tilde {\omega }}\, {\hbox {coth}}
\left( {\pi {\tilde {\omega }}\over {\kappa }}\right) 
T_{\mu\nu} [u_{\omega lm }^{up},u_{\omega lm}^{up*}]
+\int_{0}^{\infty }d\omega \, 
T_{\mu\nu}[u_{\omega lm }^{in} ,u_{\omega lm}^{in*}]\right)
\label{eq:u-}
\end{eqnarray}
\end{mathletters}
where the contribution to the stress-energy tensor, 
for a massless scalar field mode in Ricci-flat
Kerr space-time, assuming conformal coupling,
is
\begin{equation}
T_{\mu\nu}[u,u^*]=
{1\over 3}(u_{;\mu }u^*_{;\nu}+u^*_{;\mu } u_{;\nu })
-{1\over {6}}(u_{;\mu \nu }u^*+u^*_{;\mu \nu }u) 
-{1\over {6}}g_{\mu\nu}u_{;\tau }u^{*;\tau }.
\end{equation}

Kay and Wald \cite{kay-wald} have shown that there does not exist a
Hadamard state  which respects the symmetries of the space-time and
is regular everywhere in Kerr space-time.
In the absence of such a `true Hartle-Hawking' vacuum we consider the 
following states, which are attempts in the literature 
to define a thermal
state with most (but not all) of the properties of the 
Hartle-Hawking state.

The first state is that introduced by Candelas, Chrzanowski and Howard
\cite{cch}, which is constructed by thermalizing the `in' and `up'
modes with respect to their natural energy, so
\begin{eqnarray}
  \label{eq:cch_green}
 G_{CCH}(x,x') &=& \langle CCH|\hat \Phi(x) \hat \Phi(x')|CCH\rangle 
\nonumber \\
&=&  \sum_{l,m}\left(
\int_0^\infty d{\tilde\omega}\, \coth\left( 
  {\pi\tilde\omega\over \kappa}\right) 
   u_{\omega lm}^{up}(x)u_{\omega lm}^{up*}(x') 
\right. \nonumber \\ & & \left.
+ \int_{0}^{\infty }d\omega \, 
\coth \left( {\pi \omega \over \kappa}\right)
u_{\omega lm }^{in}(x)u_{\omega lm}^{in*}(x')\right) .     
\end{eqnarray}
and
\begin{eqnarray}
  \label{eq:cch}
 \langle CCH|{\hat T}_{\mu\nu}|CCH\rangle 
& = & \sum_{l,m}\left(
\int_0^\infty d{\tilde\omega}\, \coth\left( 
  {\pi\tilde\omega\over \kappa}\right) 
   T_{\mu\nu}[u_{\omega lm}^{up},u_{\omega lm}^{up*}] 
\right. \nonumber \\ & & \left.
+ \int_{0}^{\infty }d\omega \, 
\coth \left( {\pi \omega \over \kappa}\right)
T_{\mu\nu}[u_{\omega lm }^{in} ,u_{\omega lm}^{in*}]\right) .     
\end{eqnarray}
As such, it might naturally, be described as the `past Hartle-Hawking'
vacuum, however, given the discussion above it is not surprising that
as we shall show in detail below, this definition gives a state which 
does not respect the simultaneous $t$-$\phi$ reversal invariance of 
Kerr space-time.

The second state we shall consider is that introduced by Frolov and
Thorne \cite{frolov-thorne} who used the `$\eta$ formalism' to treat
the quantization of the superradiant modes.  They derived the
following expressions in the state, denoted here by $|FT\rangle$,
which they claim defined the Hartle-Hawking vacuum (at least close to
the horizon):
\begin{eqnarray}
 G_{FT}(x,x') &=& \langle FT|\eta\hat \Phi(x)\eta 
\hat \Phi(x')\eta |FT\rangle 
\nonumber\\
&=& \sum_{l,m}\left(
\int_{0}^{\infty }d{\tilde {\omega }}\, 
 \coth\left( {\pi {\tilde \omega}\over \kappa}\right) 
  u_{\omega lm}^{up}(x)u_{\omega lm}^{up*}(x') 
\right. \nonumber \\ & & \left.  
+ \int_0^\infty d\omega \, 
  \coth\left( {\pi \tilde \omega \over \kappa}\right) 
       u_{\omega lm }^{in}(x)u_{\omega lm}^{in*}(x')\right)
\label{eq:ft_green}
\end{eqnarray}
and
\begin{eqnarray}
\langle FT|{\hat {T}}_{\mu\nu}|FT\rangle 
& = & \sum_{l,m}\left(
\int_{0}^{\infty }d{\tilde {\omega }}\, 
 \coth\left( {\pi {\tilde \omega}\over \kappa}\right) 
  T_{\mu\nu}[u_{\omega lm}^{up},u_{\omega lm}^{up*}] 
\right. \nonumber \\ & & \left.  
+ \int_0^\infty d\omega \, 
  \coth\left( {\pi \tilde \omega \over \kappa}\right) 
    T_{\mu\nu}[u_{\omega lm }^{in} ,u_{\omega lm}^{in*}]\right) .
\label{eq:ft}
\end{eqnarray}
Thus, the Frolov-Thorne state differs in its choice of the appropriate
`energy' for the thermal factor corresponding to the `in' modes.  This
state is formally invariant under simultaneous $t$-$\phi$
reversal. Frolov and Thorne claim that the state defined by
Eq. (\ref{eq:ft}) is regular out to the speed-of-light surface and
is ill-defined outside. However, Kay and Wald's result is essentially
local and the Frolov-Thorne state appears to violate the spirit if not
the letter of the result proved by Kay and Wald.

Below and in subsequent papers in this series where we address the
issues numerically, we shall show that the  Frolov-Thorne state is
fundamentally flawed while the Candelas-Chrzanowski-Howard  state is
workable but cannot claim to represent an equilibrium state.

\section{Constraints on the stress tensor}
\label{sec:constraints}

We now investigate how much information can be gathered about the
stress-energy tensor in Kerr space-time from general physical
principles.  We shall have in mind the physical vacua which have been
defined in the previous section.
 
\subsection{Solution of the conservation equations}
\label{subsec:conservation}

In this section, we consider the solution of the 
conservation equations
$\nabla_{\nu }T_{\mu }{}^{\nu }=0$.  
To avoid the calculation of
Christoffel symbols, since $T_{\mu\nu }$ 
is a symmetric tensor,
the conservation equations may be written in the alternative form
\cite{dirac}
\begin{equation}
\partial_{\nu }(T_{\mu }{}^{\nu }{\sqrt {-g}}\, )
={\textstyle\frac{1}{2}} {\sqrt {-g}} \, 
(\partial_{\mu }g_{\alpha \beta })T^{\alpha \beta }
\label{eq:cons}
\end{equation}
where $g$ is the determinant of the matrix of metric 
coefficients given by
$g=-\rho^4\sin ^2\theta$.
Since we are interested in the renormalized stress tensor for states
which respect the symmetries of the space-time, we assume that 
the stress-energy tensor, like the metric, is independent of
$t$ and $\phi $. The $\mu =t$ and $\mu =\phi $ equations then become,
respectively,
\begin{eqnarray}
\partial_{r}(\rho^2 \sin \theta \, T_{t}{}^{r})
+\partial_{\theta}(\rho^2 \sin \theta \, T_{t}{}^{\theta })&=& 0
\nonumber \\ 
\partial_{r}(\rho^2 \sin \theta \, T_{\phi }{}^{r})
+\partial_{\theta }
(\rho^2 \sin \theta \, T_{\phi }{}^{\theta })&=&0.
\end{eqnarray}
These may be integrated immediately over $r$ to yield
\cite{hiscock}
\begin{eqnarray}
T_{tr}&=&
{K(\theta )\over \Delta}-{1\over {\Delta \sin \theta }}
\partial_{\theta }\left(\sin \theta \,
\int_{r_+}^{r} T_{t\theta }\>dr'\right)
\nonumber \\
T_{\phi r}&=&
{L(\theta )\over \Delta}-{1\over {\Delta\sin \theta }}   
\partial_{\theta }\left ( \sin \theta \,\int_{r_+}^{r}
T_{\phi\theta }\>dr' \right)
\label{eq:4.3}
\end{eqnarray}
where $K(\theta )$ and $L(\theta )$ are arbitrary functions of
$\theta $ alone.

The $\mu =r$ and $\mu =\theta $ equations are, respectively,
\begin{eqnarray}
F(r,\theta ) & = & 
\partial_{r}(\rho^2 T_{r}{}^{r})+\Delta ^{-1} \csc \theta \, 
\partial_{\theta }(\rho^2 \sin \theta \, T_{\theta }{}^{r})
-rT_{\theta}{}^{\theta}
\nonumber \\ & & 
-\Delta ^{-1}
(ra^2\sin \theta -\Lambda )T_{r}{}^{r}
\nonumber \\ 
G(r,\theta ) & = & 
\partial_{r}(\rho^2 T_{\theta }{}^{r}) 
+\csc \theta \, \partial_{\theta }
(\rho^2 \sin \theta \, T_{\theta}{}^{\theta})
\nonumber \\ & & 
+
a^2\sin \theta \cos \theta \,  
T_{r}{}^{r}+a^2\sin \theta \cos \theta \,
T_{\theta}{}^{\theta}
\label{eq:3.18}
\end{eqnarray}
where
\begin{eqnarray}
F(r,\theta )&=&
\rho^{-2}[-\Lambda T^{tt}+2a\Lambda \sin ^2\theta
\, T^{t\phi }+\sin ^2\theta \, (-\Lambda a^2\sin ^2\theta
+ r\rho^4)T^{\phi \phi }] 
\nonumber \\
G(r,\theta )&=&
{a^2 (r^2 + a^2 - \Delta) \over \rho^2 \Delta (r^2+a^2)} 
\sin \theta \cos \theta \, 
\left[(r^2+a^2)^2T_{tt}+2a(r^2+a^2)T_{t\phi } 
\right.
\nonumber \\ & & \left.
+ a^2 T_{\phi \phi}\right] 
+ {\rho^2 \cos\theta \over (r^2+a^2) \sin^3 \theta} T_{\phi \phi}
\end{eqnarray}
with $\Lambda =M(r^2-a^2\cos ^2\theta )$.
Here we have two equations in six unknowns each of 
which is a function of two variables $r$ and $\theta$.

One other symmetry immediately apparent from the form 
of the metric is invariance under the transformation
\begin{equation}
\theta \to {\tilde {\theta }}=\pi -\theta .
\end{equation}
The components of the stress-energy tensor will also 
possess this symmetry, so in particular
\begin{equation}
\partial_{\theta }(T_{\mu\nu})=0 
\qquad  \mathrm{when\ } \theta =\pi / 2.
\end{equation}
This does not imply that any components of
$T_{\mu\nu}$ vanish, so $T_{r\theta }$ is non-zero in general.
However, from the conservation equations (\ref{eq:3.18}), 
it follows that
\begin{equation}
T_{r\theta }=0 \qquad \mathrm{when\ } \theta =\pi /2.
\end{equation}

The other symmetry of the geometry which should be mentioned here is
invariance under simultaneous $t$-$\phi $ reversal, that is,
$t\rightarrow -t$ and $\phi \rightarrow -\phi $.  The stress tensor
for a state satisfying this invariance must have
$T_{tr}=T_{t\theta}=T_{\phi r} = T_{\phi \theta} = 0$ and
correspondingly $K(\theta)=L(\theta)=0$.  It might be thought that
this simple symmetry of the space-time should be mirrored by the
stress tensor for the physical vacua in which we are
interested. However, as discussed above this is not the case, because
of the superradiant modes. Neither the Boulware vacuum $|B^-\rangle$
nor the Unruh vacuum $|U^-\rangle$ defined in 
section \ref{subsec:vacuums}
is invariant under
simultaneous $t$-$\phi $ reversal.  This in contrast to the situation
for Schwarzschild black holes, where the Boulware vacuum is
time-reversal invariant, although the Unruh vacuum is not, due to the
Hawking flux.  In Schwarzschild space-time, the Hartle-Hawking state
is also time-reversal invariant.  Of the two Hartle-Hawking-like
states, $|CCH\rangle$ is not invariant under simultaneous $t$-$\phi $
reversal but $|FT\rangle$ is.  In section \ref{sec:asymptotics} we
shall consider further the symmetry and other properties of these
states.

\subsection{The trace anomaly}
\label{subsec:trace}

As is well-known,  conformally invariant field theories on a 
curved background $g_{\mu\nu}$ possess a conformal anomaly 
which means that the renormalized stress tensor has a
trace even though the classical stress tensor must be trace-free.
As it arises from the renormalization procedure,  the trace  
anomaly is a  geometrical scalar, depending
only on the geometry and  the nature of the quantum field under 
consideration, not on the actual quantum state.  
All methods of regularization agree that it has the form 
\begin{equation}
\langle {\hat {T}}_{\alpha}^{\alpha}\rangle_{ren}
=k_1C_{\alpha \beta \gamma \delta } 
C^{\alpha \beta \gamma \delta }
+k_2(R_{\alpha \beta }R^{\alpha \beta }-
{1\over 3}R^2)+k_3\nabla_{\alpha }\nabla ^{\alpha } R 
\end{equation}
in four dimensions.  Here $k_1$, $k_2$, $k_3$ are constants which are
independent of the space-time geometry and depend only on the quantum
field.  For example, for a massless scalar field, $k_1=k_2=k_3=(2880
\pi ^2)^{-1}$.  Although all methods of regularization agree on the
values of $k_1$, $k_2$, $k_3$ for scalar and neutrino fields, and on
$k_1$ and $k_2$ for the electromagnetic field, there is disagreement
on the value of $k_3$.  Dimensional regularization gives $k_3=0$
whilst both point separation and $\zeta $-function renormalization
give $k_3=- (96 \pi ^2)^{-1}$.  This discrepancy is unimportant for us
as $R=0$ for a Kerr black hole.  For a Kerr black hole of mass $M$ and
angular momentum $Ma$,
\begin{eqnarray}
 C_{\alpha \beta \gamma \delta }C^{\alpha \beta \gamma \delta }
& = & 48\rho^{-12}
\left\{   M^{2}r^8-15M^2r^4 a^{2} \cos ^{2} \theta   
\right. \nonumber \\ & & \left.
 + 15M^2r^2 a^{4} \cos ^{4} \theta 
-M^2a^{6}\cos ^{6}\theta \right\} 
\end{eqnarray}
where, as before, $\rho^2 =r^2+a^2\cos^2\theta$.  The trace anomaly
is, of course, finite except at a curvature singularity of the
space-time.

We may now replace one of the stress tensor components by the trace.
Hence we may substitute
\begin{equation}
T_{\theta}{}^{\theta}=T_{\alpha}{}^{\alpha}-T_{t}{}^{t}
-T_{r}{}^{r}-T_{\phi }{}^{\phi }
\end{equation}
to yield
\begin{eqnarray}
{\tilde {F}}(r,\theta ) & = & 
\partial_{r}(\rho^2 T_{r}{}^{r})+\Delta ^{-1}\csc \theta \, 
\partial_{\theta }(\rho^2 \sin \theta \, 
T_{\theta }{}^{r})+T_{r}{}^{r}
\nonumber \\ & & 
-\Delta ^{-1}(ra^2
\sin ^2\theta -\Lambda )T_{r}{}^{r}
\nonumber \\
{\tilde {G}}(r,\theta ) & = & 
\partial_{r}(\rho^2 T_{\theta }{}^{r})- \csc \theta \, 
\partial_{\theta }(\rho^2 \sin \theta \, T_{r}{}^{r})
\label{eq:4.5}
\end{eqnarray}
where
\begin{mathletters}
\begin{eqnarray}
{\tilde {F}}(r,\theta )&=&
F(r,\theta )+rT_{\alpha }{}^{\alpha }-rT_{t}{}^{t}
-rT_{\phi}{}^{\phi} 
\nonumber \\
 &=& rT_{\alpha }{}^{\alpha } + 
{(M-r) \over \Delta^2}\bigl[(r^2+a^2)^2T_{tt}
+2a(r^2+a^2)T_{t\phi } + a^2 T_{\phi \phi}\bigr] 
\nonumber \\ & & 
+
 {2r \over \Delta} \bigl[ (r^2+a^2)T_{tt}+ aT_{t\phi } \bigr] \\
{\tilde {G}}(r,\theta )&=&
G(r,\theta )-a^2\sin \theta \cos \theta \,  
(T_{\alpha }{}^{\alpha }-T_{t}{}^{t}-T_{\phi}{}^{\phi})
\nonumber \\ & & 
- \csc\theta \, \partial_{\theta }
\bigl(\rho^2 \sin \theta \, [T_{\alpha }{}^{\alpha }
-T_{t}{}^{t}-T_{\phi}{}^{\phi}]\bigr) 
\nonumber \\
&=& - {1 \over \Delta \sin \theta} 
\partial_\theta \Bigl(\sin \theta \, 
\bigl[(r^2+a^2)^2T_{tt}+2a(r^2+a^2)T_{t\phi } 
+ a^2 T_{\phi \phi}\bigr] \Bigr)
\nonumber \\ & & 
+ 2a \cot \theta (a \sin^2 \theta \, T_{tt} + T_{t\phi})
  + a^2 \sin\theta \, \partial_\theta T_{tt} 
+ 2a \partial_\theta T_{t \phi} 
\nonumber \\ & & 
+  \csc^2 \theta \, \partial_\theta T_{\phi \phi}  
-\rho^2 \partial_\theta T_\alpha{}^\alpha + \cos\theta(a^2
\sin\theta - \rho)^2 T_\alpha{}^\alpha .
\end{eqnarray}
\end{mathletters}
Equations (\ref{eq:4.5}) can be written in the
alternative form:
\begin{eqnarray}
\partial_{r}(\Delta ^{1\over 2}\rho^2 \sin \theta \, 
T_{r}{}^{r})+ \partial_{\theta }
(\Delta ^{-{1\over 2}}\rho^2 \sin \theta \, T_{\theta }{}^{r})
& = & 
\Delta ^{1\over 2}{\tilde {F}}(r,\theta )\sin \theta 
\nonumber \\
\partial_{r}(\rho^2 \sin \theta \, T_{\theta }{}^{r})
-\partial_{\theta } (\rho^2 \sin \theta \, T_{r}{}^{r})
& = & {\tilde {G}}(r,\theta )\sin \theta .
\label{eq:4.7}
\end{eqnarray}
These equations  can now be integrated over $r$ to give
\begin{eqnarray}
T_{r}{}^{r}& = & 
{R(\theta )\over {\Delta ^{1\over 2}\rho^2 }}+
{1\over {\Delta ^{1\over 2}\rho^2 \sin \theta }}
\int_{r_+}^{r} (\Delta ^{1\over 2}
{\tilde {F}}(r',\theta )\sin \theta -
\Delta ^{-{1\over 2}}\partial_{\theta }
(\rho^2 \sin  \theta \, T_{\theta }{}^{r})) dr'
\nonumber \\
T_{\theta }{}^{r} & = & 
{S(\theta )\over {\rho^2 }}+{1\over {\rho^2 \sin \theta }}
\int_{r_+}^{r}({\tilde {G}}(r',\theta )\sin \theta 
+\partial_{\theta } (\rho^2 \sin \theta \, 
T_{r}{}^{r})) dr'
\label{eq:4.9}
\end{eqnarray}
where $R(\theta )$, $S(\theta )$ are 
arbitrary functions of $\theta $ alone.
Choice of a particular vacuum state will place 
restrictions on the four arbitrary functions $K(\theta )$, 
$L(\theta )$, $R(\theta )$ and $S(\theta )$
and also on ${\tilde {F}}(r,\theta )$ and 
${\tilde {G}}(r,\theta )$, which 
depend on three unknown stress tensor components,
$T_{tt}$, $T_{t\phi }$ and $T_{\phi \phi }$.
The solutions (\ref{eq:4.9}) are particularly useful for 
finding the behaviour of the stress tensor close to the
event horizon, but we still have the coupling between 
$T_{r}{}^{r}$ and $T_{\theta }{}^{r}$.

Uncoupled equations for $T_{r}{}^{r}$ and $T_{\theta }{}^{r}$,
can be obtained from (\ref{eq:4.7}) in the form:
\begin{eqnarray}
\Delta ^{\frac {1}{2}} \partial _{r} \left[
\Delta ^{\frac {1}{2}} \partial _{r} {\mathcal{T}}_1 \right] +
\partial ^{2}_{\theta } {\mathcal{T}}_1 
& = & 
\Delta ^{\frac {1}{2}} \partial _{r} \left(
{\mathcal{F}} \right) -\partial _{\theta } \left(
{\mathcal{G}} \right)
\nonumber \\
\Delta ^{\frac {1}{2}} \partial _{r} \left[
\Delta ^{\frac {1}{2}} \partial _{r} \left(
{\mathcal{T}}_2 \right) \right] +
\partial ^{2}_{\theta } \left( {\mathcal{T}}_2 \right)
& = & 
\Delta ^{\frac {1}{2}} \partial _{r} \left(
{\mathcal{G}} \right) +
\partial _{\theta } \left( {\mathcal{F}} \right) 
\label{eq:elliptic}
\end{eqnarray}
where
\begin{equation}
{\mathcal{T}}_1=T_{r}{}^{r}
\Delta ^{\frac {1}{2}} \rho ^{2} \sin \theta 
\qquad
{\mathcal{T}}_2=T_{\theta }{}^{r}
\rho ^{2} \sin \theta 
\qquad
{\mathcal{F}}=\Delta {\tilde {F}} \sin \theta
\qquad
{\mathcal{G}}=\Delta ^{\frac {1}{2}} {\tilde {G}} \sin \theta .
\label{eq:calt}
\end{equation}
We now define a new variable $x$ by:
\begin{equation}
x=2\Delta ^{\frac {1}{2}} +2r-2M, 
\end{equation} 
in terms of which the equations (\ref{eq:elliptic}) 
now have the usual polar form of the Laplacian:
\begin{mathletters}
\begin{eqnarray}
x\partial _{x} \left[ x\partial _{x} {\mathcal{T}}_1
\right] +
\partial ^{2}_{\theta } {\mathcal{T}}_1 
& = & 
x\partial _{x} {\mathcal{F}} -\partial _{\theta } {\mathcal{G}}
\label{eq:polar_a} \\
x\partial _{x} \left[ x\partial _{x} {\mathcal{T}}_2
\right] +
\partial ^{2}_{\theta } {\mathcal{T}}_2
& = & 
x\partial _{x} {\mathcal{G}} +\partial _{\theta } {\mathcal{F}} .
\end{eqnarray}
\label{eq:polar}
\end{mathletters}
The domain of these equations is 
$x\in ((r^{2}-a^{2})^{\frac {1}{2}},\infty )$,
$\theta \in (0,\pi )$, that is the punctured half-plane.
By constructing a Green's function for this domain, a unique 
solution for ${\mathcal{T}}_1$ and ${\mathcal{T}}_2$ can be found
if they are specified on the boundary, 
provided we know $\mathcal{F}$ and $\mathcal{G}$ throughout
the region.
Therefore, we need to know $T_{r}{}^{r}$ and $T_{\theta }{}^{r}$
on the event horizon (where $x=(r^{2}-a^{2})^{\frac {1}{2}}$),
and the three components of the stress tensor, $T_{tt}$,
$T_{t\phi }$ and $T_{\phi \phi }$ everywhere outside the event
horizon. From Eq. (\ref{eq:calt}), it can be seen that ${\mathcal{T}}_1$
and ${\mathcal{T}}_2$ must vanish on the axis $\theta =0,\pi $
provided that $T_{r}{}^{r}$ and $T_{\theta }{}^{r}$ are
well-defined there.
Therefore this reduces the number of boundary functions which
are unknown.

Although it looks like ${\mathcal{T}}_1$ vanishes on the event
horizon, the analysis of subsection \ref{subsec:hor}
will show that even for a quantum state which is regular
on the event horizon, $T_{r}{}^{r}$ diverges as $\Delta ^{-1}$
as $r\rightarrow r_{+}$, giving a divergent value
for ${\mathcal{T}}_1$ on the horizon. 
This means that the Green's function method is not directly applicable
to Eq. (\ref{eq:polar_a}).
However, the second equation can be solved uniquely using a
Green's function, and the solution then fed into 
Eq. (\ref{eq:4.9}) to give the behaviour of $T_{r}{}^{r}$. 
Note that our calculations in section \ref{sec:asymptotics}
confirm that, for the Unruh and (past and future) Boulware vacua, 
the function ${\mathcal{T}}_2$ vanishes sufficiently quickly at infinity
that the Green's function method gives a unique solution.

\subsection{The Killing-Yano Tensor}
\label{subsec:killing_yano}

So far in our analysis we have exploited the Killing vector 
symmetries of the Kerr geometry to assume that the stress 
tensor is a function only of $r$ and $\theta $.
The Kerr geometry also possesses a Killing-Yano tensor 
\cite{walker-penrose}, which is a
skew-symmetric tensor $f_{\mu \nu }$ satisfying
\begin{equation}
\nabla ^{(\mu }f^{\nu )\lambda }=0.
\end{equation}
We shall now show that the consequence of the existence of the
Killing-Yano tensor is that $T_{x \theta }=0$, when $x=t$ or 
$x=\phi $, for the quantum states we are interested in.

For any quantum state, the renormalized expectation value of the
quantum stress tensor can be calculated using the technique of point
splitting:
\begin{equation}
\langle T_{\mu \nu } \rangle _{{\rm {ren}}} =
\lim _{x \rightarrow x'} \left[
T_{\mu \nu }(x,x') -T_{\mu \nu }^{{\rm {div}}} (x,x') \right]
\end{equation}
where $T_{\mu \nu }(x,x')$ is the point-separated stress tensor for
our particular quantum state and $T_{\mu \nu }^{{\rm {div}}}(x,x')$
are the divergent subtraction terms.
The unrenormalized stress tensor components for the quantum states in
which we are interested are given as mode sums (\ref{eq:b-}--\ref{eq:ft}), 
the mode sum contribution to $T_{A\theta }$
for $A=t$ or $A=\phi $ being
\begin{equation}
T_{A\theta }[u,u^*]=\Re e \left[
\frac {1}{3} \left( u_{;A}u^{*}_{;\theta } +u^{*}_{;A}u_{;\theta }
\right)
-\frac {1}{6} \left( u_{;A\theta }u^{*} +u^{*}_{;A\theta }u \right)
\right] .
\label{eq:xtheta}
\end{equation}
The existence of the Killing-Yano tensor has the result that the wave
equation for a massless scalar field on the Kerr geometry is
separable \cite{carter}, 
with the mode solutions given by (\ref{eq:modes}).
In addition, we have
\begin{equation}
u_{;A\theta }=u_{,A\theta }-\Gamma ^{t}_{A\theta }u_{,t}
-\Gamma ^{\phi }_{A\theta }u_{,\phi } .
\end{equation}
From the mode functions, $u_{,A}=iku$ where $k=-\omega $ if $A=t$ and
$k=m$ if $A=\phi $; also
\begin{equation}
u_{,\theta }\propto -\left( r^{2}+a^{2} \right) ^{-\frac {1}{2}}
e^{-i\omega t+im\phi } R_{\omega lm}(r) S'_{\omega lm }(\cos \theta )
\sin \theta  ;
\end{equation}
and $u_{,x\theta }=iku_{,\theta }$.
Since the spheroidal harmonics $S_{\omega lm}$ are real, the
quantities appearing in (\ref{eq:xtheta}) are all purely imaginary and
hence $T_{x\theta }[u,u^{*}]=0$.
Therefore the point-separated stress tensor components 
$T_{A\theta }(x,x')$
vanish for all of the states under consideration.
In addition, it is shown in \cite{frolov-thorne} that 
the subtraction terms 
$T_{A\theta }^{{\rm {div}}}(x,x')$ are also zero, so that 
$\langle T_{x\theta } \rangle _{{\rm {ren}}}$ vanishes 
for all the states
we are considering here. This property was proved by Frolov and Thorne
\cite{frolov-thorne} for $|FT\rangle$ but we have shown here that
this is a quite general result.

\subsection{Behaviour on the event and Cauchy horizons}
\label{subsec:hor}

Next we shall investigate the behaviour of the stress 
tensor at the future and
past event and Cauchy horizons.  
It is convenient to introduce co-ordinate systems
that are regular at the horizons.  
We first introduce two double-null co-ordinate systems 
$u$, $v$, $\theta$, $\phi_\pm$ by
\begin{equation}
u=t-r_*, \quad v=t+r_*, \quad \phi_\pm=\phi -{a\over {r_\pm^2+a^2}}
t=\phi -\Omega_\pm t,
\end{equation}
where the last equation defines $\Omega_+$ and $\Omega_-$ which are
the angular velocity of the event and Cauchy horizons respectively,
and $r_{*}$ is the `tortoise' co-ordinate given by Eq. (\ref{eq:tortoise}).
The two sets of Kruskal co-ordinates $U_\pm$, $V_\pm$ are then defined by
\begin{equation}
U_\pm =-e^{-\kappa_\pm u}, \quad V_\pm =e^{\kappa_\pm v} .
\end{equation}
From the definition of $U_\pm$, $V_\pm$ and $r_*$,
\begin{equation}
U_\pm V_\pm =-e^{2\kappa_\pm r_*}=-
e^{2\kappa_\pm r}(r-r_\pm)|r-r_\mp|^
{\kappa_\pm / {\kappa_\mp}}.
\end{equation}
The exterior region corresponds to $U_+<0$, $V_+>0$ with the past
event horizon at $V_+=0$ and the future event horizon at $U_+=0$.
These coordinates may be extended to cover the event
horizons in a regular fashion but are singular at the Cauchy 
horizons.
Correspondingly, the coordinates $U_-$ and $V_-$ may be extended to
cover the Cauchy horizons ($U_-=0$ and $V_-=0$) in a regular fashion 
but are singular at the event horizons.

The stress tensor components in these Kruskal co-ordinate systems
are
\begin{mathletters}
\label{eq:kruskal}
\begin{eqnarray}
T_{U_\pm  U_\pm}&=&\kappa_\pm^{-2}U_\pm^{-2}
\left[ {1\over 4}T_{tt}+{1\over 2}\Omega_\pm T_{t\phi }+{1\over 4}
\Omega_\pm^2T_{\phi \phi }-{\Delta \over {2(r^2+a^2)}}T_{rt}
\right. \nonumber \\ & & \left.
+ {\Delta ^2\over {4(r^2+a^2)^2}}T_{rr}
-{\Delta \Omega_\pm\over  {2(r^2+a^2)}}T_{r\phi}
     \right]
\\
T_{U_\pm  V_\pm}&=&- \kappa_\pm^{-2}U_\pm^{-1}V_\pm^{-1}
\biggl[{1\over 4}T_{tt}+{1\over 2}\Omega_\pm T_{t\phi }+{1\over 4}
\Omega_\pm^2 T_{\phi \phi }- {\Delta ^2\over {4(r^2+a^2)}}T_{rr} \biggr]
\\
T_{V_\pm  V_\pm}&=&\kappa_\pm^{-2}V_\pm^{-2}
\left[ {1\over 4}T_{tt}+{1\over 2}\Omega_\pm T_{t\phi }+{1\over 4}
\Omega_\pm^2 T_{\phi \phi } + {\Delta \over {2(r^2+a^2)}}T_{rt}
\right. \nonumber \\ & & \left.
+ {\Delta ^2\over {4(r^2+a^2)^2}}T_{rr} + 
{\Delta \Omega_\pm \over {2(r^2+a^2)}}T_{r\phi} \right]
\\
T_{U_\pm \theta }&=&\kappa_\pm^{-1}U_\pm^{-1}
{\Delta \over {2(r^2+a^2)}}T_{r\theta } 
\\
T_{V_\pm \theta }&=&\kappa_\pm^{-1}V_\pm^{-1}
{\Delta \over {2(r^2+a^2)}}T_{r\theta } 
\\
T_{U_\pm\phi_\pm}&=&-\kappa_\pm^{-1}U_\pm^{-1}
\biggl[{1\over 2}T_{t\phi }+{1\over 2}\Omega_\pm T_{\phi \phi }
-{\Delta \over {2(r^2+a^2)}}T_{r\phi  }\biggr] 
\\
T_{V_\pm\phi_\pm}&=&\kappa_\pm^{-1}V_\pm^{-1}
\biggl[{1\over 2}T_{t\phi }+{1\over 2}\Omega_\pm T_{\phi \phi }
+{\Delta \over {2(r^2+a^2)}}T_{r\phi  }\biggr] 
\end{eqnarray}
\end{mathletters}
with $T_{\theta \theta }=T_{\theta \theta }$, 
$T_{\theta  \phi_\pm}=0$ 
and $T_{\phi_\pm \phi_\pm} = T_{\phi \phi }$, where we have set
$T_{t\theta }=T_{\theta \phi }=0$. 
It follows immediately that regularity of the stress tensor on any
horizon
requires that $T_{\theta \theta}$,   $T_{\phi \phi }$
and $T_{r \theta}$ be finite as the horizon is approached.  

For a general stress tensor with $T_{t \theta} = T_{\theta \phi} = 0$,
we have by Eqs. (\ref{eq:4.3})
\begin{equation}
  T_{tr} = {K(\theta) \over \Delta}, \qquad 
  T_{\phi r } = {L(\theta) \over \Delta} .
\label{eq:kl}
\end{equation}
In this case, consideration of the 
$T_{U_\pm \phi_\pm}$ and  $T_{V_\pm \phi_\pm}$
components shows that regularity requires
\begin{equation}
  T_{t \phi}(r,\theta) = \pm {L(\theta) \over r_\pm^2 + a^2} 
- \Omega_\pm
  T_{\phi\phi}(r_\pm,\theta) + O(r- r_\pm) 
\qquad {\mathrm{as\  }} r \to r_\pm,
\label{eq:ttphor}
\end{equation}
where the positive sign is taken for regularity on the
future horizon ($U_{\pm }=0$) and the negative
sign on the past horizon ($V_{\pm }=0$).
Note that if $L(\theta)$ is non-zero, only one of these 
conditions can be met on either the  future or past event horizon.
Regularity of the  $T_{U_{\pm }U_{\pm}}$, $T_{V_{\pm }V_{\pm }}$ 
and $T_{U_{\pm }V_{\pm }}$ components implies that
\begin{mathletters}
\begin{eqnarray}
T_{tt} & = &
\pm 
\frac {K(\theta )-\Omega _{\pm }L(\theta )}{r^{2}+a^{2}} 
+\Omega _{\pm }^{2}T_{\phi \phi }+
O(r-r_{\pm }) 
\label{eq:ttthor}
\\
T_{rr} & = & 
\pm \left[ K(\theta )+\Omega _{\pm }L(\theta ) \right]
\frac {r^{2}+a^{2}}{\Delta ^{2}} +O(r-r_{\pm })^{-1}
\label{eq:trrhor}
\end{eqnarray}
\end{mathletters}
as $r\to r_{\pm }$,
with the positive sign for regularity on the future horizon, and
the negative sign for the past horizon, as before.
Finiteness of $T_{r\theta }$ as the horizon is approached implies that
the function $S(\theta )$ in (\ref{eq:4.9}) vanishes, whilst the form
(\ref{eq:trrhor}) of $T_{rr}$ near the horizon tells us that 
$R(\theta )$ in (\ref{eq:4.9}) is also identically zero.
It should be stressed that the forms 
(\ref{eq:ttphor}-\ref{eq:trrhor}) are compatible with the
solution of the conservation equations (\ref{eq:4.9}) with
$R$ and $S$ identically zero.

We note that our analysis is in agreement with that of
\cite{hiscock}, in that unless both $K(\theta)$ and $L(\theta)$ 
vanish identically,
the stress tensor must diverge at one of the event horizons,
and at least one of the Cauchy horizons.
The past and future Boulware vacua are not expected 
to be regular on either 
event horizon.
For the Unruh vacuum state, it is expected that the divergences 
occur on the past event horizon and future Cauchy horizon 
\cite{hiscock}.
For $|FT\rangle$ simultaneous $t$-$\phi$ invariance required 
that $K(\theta)$ and $L(\theta)$ vanish consistent with regularity.
On the other hand, for $|CCH\rangle$ there was no requirement that 
 $K(\theta)$ and $L(\theta)$ vanish and so one expects that
there will be divergences on the past event horizon in line with the
Unruh vacuum.
We shall return to this issue in section \ref{sec:asymptotics}.

At this stage, we need to step back and see how much information
about the stress tensor we have managed to obtain from our
approach. 
We began with ten stress tensor components, each a function of the
two variables $r$ and $\theta $.
The Killing-Yano symmetry revealed that two of these components
$T_{t\theta }$ and $T_{\theta \phi }$ vanished identically,
whilst another component could be eliminated by using the known
trace anomaly, leaving seven unknown functions of $r$ and
$\theta $.
Using the conservation equations, we need to know three
functions of $r$ and $\theta $ (corresponding to $T_{tt}$,
$T_{t\phi }$ and $T_{\phi \phi }$), and four functions
of $\theta $ ($K$, $L$, $T_{rr}$
 and $T_{r\theta }$ on the event horizon).
Finally, for a state which is regular on one of the event horizons,
this reduces to three functions of $\theta $ since
the behaviour of $T_{rr}$ is given in terms of $K$ and $L$.
In addition, we know the behaviour of the three unknown components,
$T_{tt}$, $T_{t\phi }$ and $T_{\phi \phi }$ on the event horizon,
in terms of $K$, $L$ and $T_{\phi \phi }$.
Thus our analysis has significantly reduced the number of degrees 
of freedom of the stress tensor in Kerr space-time.
Of course, this reduction is rather less significant than the 
corresponding analysis for Schwarzschild black holes
\cite{cf}, but this was to be expected due to the fact that
Kerr has fewer symmetries than Schwarzschild.

\section{Asymptotic behaviour of the physical vacua}
\label{sec:asymptotics}

In this section we shall consider the asymptotic behaviour of the
physical states of interest near the event horizon and at infinity.
This will provide a consistency check on the analysis of the previous
section.  We shall also use the properties of the Unruh and Boulware
vacua (whose asymptotic behaviour are well understood) to reveal
information about the states $|CCH\rangle$ and $|FT\rangle$.  It is
known that the divergent terms which have to be subtracted from the
unrenormalized expectation value of the stress tensor are independent
of the quantum state under consideration.  Therefore we shall consider
the differences in expectation values of the stress tensor in two
different states, since these can be calculated without
renormalization.  Such differences in expectation values will be
traceless tensors since the trace anomaly is the same for all quantum
states.

We shall begin by concentrating on the Unruh vacuum, since 
its stress tensor has been calculated in the asymptotic regimes by
Punsley \cite{punsley} using an equivalence principle approach.
This will provide a useful check of our calculations.
Firstly, we consider the behaviour at infinity, and calculate
\begin{eqnarray}
\langle U^{-}|{\hat {T}}_{\mu\nu}|U^{-}\rangle_{ren}-
\langle B^{-}|{\hat {T}}_{\mu\nu}|B^{-}\rangle_{ren} &=&
\langle U^{-}|{\hat {T}}_{\mu\nu}|U^{-}\rangle -
\langle B^{-}|{\hat {T}}_{\mu\nu}|B^{-}\rangle 
\nonumber \\
{} &=&\sum_{l,m }\int_{0}^{\infty }
{2d\, {\tilde {\omega }}\over {e^{2\pi 
{\tilde {\omega }}/\kappa }-1}}T_{\mu\nu}
[u_{\omega lm}^{up},u_{\omega lm}^{up*}]. 
\end{eqnarray}
Using the asymptotic form of the mode functions (\ref{eq:2.3}),
we have, as $r\rightarrow \infty $,
\begin{eqnarray}
& & 
\langle U^{-}|{\hat {T}}_{\mu }^{\nu }|U^{-}\rangle _{ren}-
\langle B^{-}|{\hat {T}}_{\mu }^{\nu }|B^{-}\rangle _{ren}
\nonumber \\ 
& \sim & 
\frac {1}{4\pi ^{2}r^{2}}
\sum_{l,m}\int_{0}^{\infty } 
{\omega \,\mathrm{d}{\tilde {\omega }}\over {{\tilde {\omega }}
(e^{2\pi {\tilde {\omega }}/\kappa }-1)}}
|B_{\omega lm}^-|^2
|S_{\omega lm}(\cos \theta )|^2 
\left(
\begin{array}{cccc}
-\omega  &\omega  & 0 & m \\
-\omega  &\omega  & 0 & m \\
0 & 0 & 0 & 0 \\
0 & 0 & 0 & 0  
\end{array}   \right) .
\end{eqnarray}
In order to obtain the behaviour of the Unruh vacuum at future null 
infinity, we need to consider the `past' Boulware vacuum at infinity.
The `past' Boulware vacuum contains at future null infinity an outward
flux of
particles due to the Unruh-Starobinskii effect \cite{unruh},
so that, as we approach $\mathfrak{I}^+$,
\begin{eqnarray}
\langle B^{-}|{\hat T}_{\mu }^{\nu }|B^{-}\rangle_{ren} 
& \sim & \langle B^{-}|{\hat {T}}_{\mu }^{\nu }|B^{-}\rangle
-      \langle B^{+}|{\hat {T}}_{\mu }^{\nu }|B^{+}\rangle 
\nonumber \\ 
& \hspace{-37pt}
 \sim & \hspace{-20pt}
{1\over {4\pi ^2r^2}}\sum_{l,m}\int_{0}^{\omega_{min}}
\frac {\omega  \, d\omega }{{\tilde {\omega }}
(e^{2\pi {\tilde {\omega }}/\kappa }-1) }
|B_{\omega lm}^-|^2|S_{\omega lm}(\cos \theta )|^2 
\left( \begin{array}{cccc}
\omega & -\omega & 0 & -m \\
\omega  & -\omega & 0 & -m \\
0 & 0 & 0 & 0 \\
0 & 0 & 0 & 0  
\end{array} \right) .
\label{eq:binf}
\end{eqnarray}
Adding these two tensors gives the asymptotic 
behaviour of the Unruh vacuum at future null infinity as:
\begin{eqnarray}
\langle U|{\hat {T}}_{\mu }^{\nu }|U\rangle_{ren}
& \sim & 
{1\over {4\pi ^2r^2}}\sum_{l,m} \int_{0}^{\infty }
\frac {\omega  \, d\omega }{
{\tilde {\omega }}(e^{2\pi {\tilde {\omega }}/\kappa }-1)}
|B_{\omega lm}^-|^2 |S_{\omega lm}(\cos \theta )|^2 
\left( \begin{array}{cccc}
-\omega & \omega  & 0 & m \\
 -\omega & \omega & 0 & m \\
0 & 0 & 0 & 0 \\
0 & 0 & 0 & 0 \\
\end{array} \right) .
\label{eq:uninf}
\end{eqnarray}
This is in agreement with the form obtained in \cite{punsley},
and represents the expected thermal flux at infinity.
It should be noted that, despite initial appearances, the integrands 
are regular when ${\tilde {\omega }}=0$ due to the Wronskian
relations (\ref{eq:2.4}) which ensure that 
$|B_{\omega lm}^{-}|^{2}=O({\tilde {\omega }}^{2})$ 
as ${\tilde {\omega }}\rightarrow 0$.
From Eq. (\ref{eq:uninf}) we can read off the forms of the
functions $K$ and $L$ (\ref{eq:kl}) for the Unruh vacuum:
\begin{eqnarray}
K_{U^-}(\theta ) & = & 
\frac {1}{4\pi ^{2}} \sum _{l,m} 
\int _{0}^{\infty } 
\frac {- \omega ^{2} \, d\omega}{ {\tilde {\omega }}
(e^{2\pi {\tilde {\omega }}/\kappa }-1)} 
|B^{-}_{\omega lm} |^{2}
|S_{\omega lm}(\cos \theta )|^{2} 
\nonumber \\
L_{U^-}(\theta ) & = & 
\frac {1}{4\pi ^{2}} \sum _{l,m}
\int _{0}^{\infty } 
\frac {-m\omega  \, d\omega }{{\tilde {\omega }}
(e^{2\pi {\tilde {\omega }}/\kappa }-1)}
|B^{-}_{\omega lm}|^{2} 
|S_{\omega lm}(\cos \theta )|^{2} .
\label{eq:klu}
\end{eqnarray}

We now turn to the behaviour of the Unruh vacuum at the event
horizon.
In Schwarzschild, the Hartle-Hawking state is regular on both event
horizons, and so the behaviour of the Unruh vacuum as 
$r\rightarrow r_{+}$ is found from:
\begin{eqnarray} 
\langle U^{-}|{\hat {T}}_{\mu\nu}|U^{-}\rangle_{ren}
 & \sim  &
\langle U^{-}|{\hat {T}}_{\mu\nu}|U^{-}\rangle_{ren}-
\langle H|{\hat {T}}_{\mu\nu}|H\rangle_{ren}
\nonumber \\ &= &
\langle U^{-}|{\hat {T}}_{\mu\nu}|U^{-}\rangle -
\langle H|{\hat {T}}_{\mu\nu}|H\rangle  .
\end{eqnarray}
In the absence of a Hartle-Hawking state for Kerr,
we shall instead consider the differences of the stress tensors
in the Unruh vacuum and the states $|FT\rangle $ and $|CCH\rangle $.
These  are given by:
\begin{mathletters}
\begin{eqnarray}
\langle U^{-}|{\hat {T}}_{\mu\nu}|U^{-}\rangle -
\langle FT|{\hat {T}}_{\mu\nu}|FT\rangle
& = & 
\sum_{l,m}\int_{0}^{\infty } 
\frac {-2 \, d\omega}{e^{2\pi {\tilde {\omega }}/\kappa }-1}
T_{\mu\nu}[u_{\omega lm}^{in},u_{\omega lm}^{in*}], 
\label{eq:u-fthor} \\
\langle U^{-}|{\hat {T}}_{\mu\nu}|U^{-}\rangle -
\langle CCH|{\hat {T}}_{\mu\nu}|CCH\rangle 
& = & 
\sum_{l,m}\int_{0}^{\infty } 
\frac {-2 \, d\omega}{e^{2\pi \omega /\kappa }-1}
T_{\mu\nu}[u_{\omega lm}^{in},u_{\omega lm}^{in*}].
\end{eqnarray}
\end{mathletters}
As $r\rightarrow r_{+}$, one finds
\begin{eqnarray}
 & & 
\langle U^{-}|{\hat {T}}_{\mu }^{\nu }|U^{-}\rangle
-\langle FT|{\hat {T}}_{\mu }^{\nu }|FT \rangle
\nonumber \\
 & & 
\sim 
{1\over {4\pi ^2\rho^2 }}\sum_{l,m}\int_{0}^{\infty }
{d\omega \over {\omega (e^{2\pi {\tilde
{\omega }}/\kappa }-1)}}|B_{\omega lm}^+|^2
|S_{\omega lm}(\cos \theta )|^2
\nonumber \\
 & & 
\! \! \! \! \! \! \times \left( \begin{array}{cccc}
\Delta ^{-1}(r_+^2+a^2)\omega {\tilde {\omega }} &
-\omega {\tilde {\omega }} & 0 &  
\Delta ^{-1}a\omega {\tilde {\omega }} \\     
\Delta ^{-2}(r_+^2+a^2)^2{\tilde {\omega }}^2 & 
-\Delta ^{-1}(r_+^2+a^2)^2{\tilde {\omega }}^2 &
O(1) &    
 -\Delta ^{-2}a(r_+^2+a^2){\tilde {\omega }}^2 \\
0 & O(\Delta ) & O(1) & 0 \\ 
\Delta ^{-1}(r_+^2+a^2)m{\tilde {\omega }} & m{\tilde {\omega }} & 0 & 
-\Delta ^{-1}am {\tilde {\omega }}  
\end{array} \right) .
\label{eq:u-ft} 
\end{eqnarray}
The expression for
$\langle U^{-} |{\hat {T}}_{\mu }^{\nu } |U^{-} \rangle
-\langle CCH |{\hat {T}}_{\mu }^{\nu }|CCH \rangle $
is identical to  Eq. (\ref{eq:u-ft}), with the denominator
$e^{2\pi {\tilde {\omega }}/\kappa }-1$ 
replaced by $e^{2\pi \omega /\kappa }-1$.
In both cases the integrand is regular for all values of $\omega $,
by virtue of the Wronskian relations (\ref{eq:2.4}).
The difference in expectation values of the stress tensor in the Unruh
and Frolov-Thorne states (\ref{eq:u-ft}) agrees with the 
stress tensor for the Unruh vacuum found in \cite{punsley},
whereas when we have the state $|CCH\rangle $ instead of $|FT \rangle $
the thermal terms in the denominator do not agree.
Furthermore, the tensor (\ref{eq:u-ft}) is regular on the future event 
horizon but not on the past event horizon, the same
behaviour that we would expect for the Unruh vacuum.
Therefore we can compare the tensor (\ref{eq:u-ft})
with the behaviour near the event horizon derived
in section \ref{subsec:hor}.
There is exact agreement, using the functions
$K_{U^-}(\theta)$ and $L_{U^-}(\theta)$ found from the expectation value
of the stress tensor at infinity in the Unruh vacuum
(\ref{eq:klu}), and the Wronskian relations.

From the regularity of the tensor (\ref{eq:u-ft}) on the
future event horizon, we can conclude that the expectation value of the
stress tensor in the state $|FT\rangle $ is regular
on at least one event horizon (and, since it is invariant under
simultaneous $t$, $\phi $ reversal, it will be regular on 
both event horizons).
Thus, it may appear that the state $|FT\rangle $ in fact has the 
properties that we require of the Hartle-Hawking state.
However, whilst the expectation value of the stress tensor in the
state $|FT\rangle $ is regular on the event horizon, the
expectation value of ${\hat \Phi}^2$ is not.
We calculate, as $r\rightarrow r_{+} $,
\begin{eqnarray}
& & 
\langle U^{-} |{\hat {\Phi }}^{2}|U^{-} \rangle
-\langle FT |{\hat {\Phi }}^{2}|FT \rangle
\nonumber \\
& \sim &
\frac {1}{4\pi ^{2}(r_{+}^{2}+a^{2})}
\sum _{l,m} \int _{0}^{\infty }
\frac {-2 \, d\omega }{\omega (e^{2\pi {\tilde {\omega }}/\kappa }-1)}
|B_{\omega lm}^{+}|^{2} |S_{\omega lm}(\cos \theta )|^{2} .
\label{eq:phi2}
\end{eqnarray}
The integrand in the above expression is regular at $\omega =0$
because of the Wronskian relations (\ref{eq:2.4}), but
has a pole at $\tilde \omega=0$, giving a divergent integral. 
If we attempt to calculate the difference in expectation values
(\ref{eq:u-fthor}) anywhere outside the event horizon, then
the integral over $\omega $ also has a pole at
$\tilde \omega =0$, leading to a divergent result.
Therefore it seems that the regularity of the difference in 
expectation values of the stress tensor (\ref{eq:u-ft}) at the 
event horizon does not reflect the true nature of \emph{the state}
$|FT\rangle $, and that this state \emph{in fact fails to be regular
almost everywhere}, both on or outside the event horizon, although
it formally has attractive symmetry properties.

There is one exception to the regularity of the state $|FT\rangle $
which is that on the axis the terms with $m \neq 0$ (and, in
particular, all superradiant modes) do not contribute.  Thus, if
one point is on the axis the $|FT\rangle $ and 
$|CCH \rangle $ two-point functions
agree: 
\begin{eqnarray}
  G^{FT/CCH}(t,r,\theta,\phi;t',r',0,\phi') 
& = &
  \sum_l \int_0^\infty {d \omega \coth(\pi \omega/\kappa) \over 
       \omega\sqrt{(r^2+a^2)({r'}^2+a^2)}} 
\nonumber \\ & & 
\hspace{-15pt} \times
\left[ R^+_{\omega l0}(r)
  R^{+*}_{\omega l0}(r') +  R^-_{\omega l0}(r)  R^{-*}_{\omega l0}(r')
   \right] S_{\omega l 0}(\cos \theta) S_{\omega l 0}(1) .
\label{eq:on-axis}
\end{eqnarray}
In the asymptotic regions, the integrals are dominated by the
contribution from near $\omega=0$.  In this limit the spheroidal
functions reduce to Legendre polynomials
\begin{equation}
   S_{0lm} = {1 \over \sqrt{4\pi}} P_l(\cos \theta), \qquad \lambda(0)
   = l(l+1) .
\end{equation}
In addition, $T_{lm}(r) = R_{0 lm}(r)/\sqrt{r^2+a^2}$ satisfies the
equation
\begin{equation}
  \label{eq:omega=0}
  {\mathrm{d } \over \mathrm{d} \eta} (\eta^2-1) 
{{\mathrm{d}}T_{lm} \over
  \mathrm{d} \eta} - \left[ l(l+1) 
+ {m^2a^2 \over (M^2 - a^2)(\eta^2  -1)}
\right] T_{lm} = 0
\end{equation}
where 
\begin{equation}
\eta = {2r - (r_+ + r_-) \over (r_+ - r_-)} 
= {r-M \over \sqrt{M^2 -a^2}} ,
\end{equation}
with solutions $P_l^{ma/\sqrt{M^2-a^2}}(\eta)$ and  
$Q_l^{ma/\sqrt{M^2-a^2}}(\eta)$. 
In particular, a steepest descent analysis of Eq.\ (\ref{eq:on-axis})
as $r' \to r_+$ yields 
\begin{eqnarray}
  \label{eq:on-pole}
  G^{FT/CCH}(t,r,\theta,\phi;t',r_+,0,\phi') &=&
  {\kappa_+ \over 16\pi^2 {\sqrt {M^2-a^2}}} 
\sum_l (2l+1) Q_l\left({r-M
  \over \sqrt{M^2 -a^2}}\right) 
   P_l(\cos \theta) 
\nonumber \\
 &=& {\kappa_+ \over 8\pi^2} 
{ 1 \over r- M - \sqrt{M^2 -a^2} \cos  \theta } ,
\end{eqnarray}
where the second line follows from Heine's formula.  This result was
first given by Frolov \cite{frolov82} and enabled him to calculate the
renormalized value of the expectation value of $\hat \Phi^2$ on the
pole of the event horizon. Later with Zel'nikov~\cite{fz} he extended
this calculation to calculate the renormalized value of the
expectation value of $\hat T_{\mu\nu}$ on the pole of the event
horizon.  Our point is that, unfortunately, these calculations were
only possible because the troublesome superradiant modes do not
contribute on the axis and have actually led to a false confidence
concerning the Hartle-Hawking vacuum.

Finally we return the properties of the state $|CCH\rangle $.
This has a different thermal factor from $|FT \rangle $
(\ref{eq:cch}) which means that the difference in
expectation values of the stress tensor in $|U^{-}\rangle $
and $|CCH \rangle $ at the event horizon is rather 
different from simply the stress tensor in the state $|U^{-}\rangle $.
The difference in thermal factors also means that the state
$|CCH\rangle $ is \emph{not} invariant under simultaneous 
$t$-$\phi $ reversal.

However, the quantity 
$\langle U^{-}|{\hat {T}}_{\mu }^{\nu }|U^{-}\rangle
-\langle CCH |{\hat {T}}_{\mu }^{\nu }|CCH \rangle $
is regular on the future event horizon (but not on the past), so,
using the expected regularity of the Unruh vacuum, we can 
conclude that $\langle CCH |{\hat {T}}_{\mu }^{\nu }|CCH \rangle $
is also regular on the future event horizon (but not on the past).
If we consider the difference in expectation values of 
${\hat {\Phi }}^{2}$ at the event horizon,
the answer is the same as
(\ref{eq:phi2}), but with $e^{2\pi {\tilde {\omega }}/\kappa }$
replaced by $e^{2\pi \omega /\kappa }$.
Using the Wronskian relations (\ref{eq:2.4}), this gives 
a finite answer, further strengthening our argument
that $|CCH\rangle $ is a regular state on the future event
horizon.

\section{Conclusions}
In this paper we have considered the renormalized stress
energy tensor on Kerr space-time, and used the anticipated 
physical properties of this tensor (symmetry, conservation
equations, and regularity conditions) in order to 
derive as much information as possible.
As expected, the analysis is considerably more complex
than the corresponding problem in Schwarzschild 
\cite{cf}, and the solution gives us less information,
although we are able to reduce the number of unknowns
to three functions of $r$ and $\theta $ and three
functions of $\theta $.

Our results are in agreement with the known form
of the Unruh vacuum at the event horizon and at infinity.
We also considered two candidates for the state
analogous to the Hartle-Hawking state in Schwarzschild. From 
the Kay-Wald theorem \cite{kay-wald}, we know that
there is no state in Kerr which is regular at the event 
horizon and everywhere outside, invariant under 
simultaneous $t$, $\phi $ reversal and thermal in nature.
Of our two candidate states, one is invariant under
$t$, $\phi $ reversal, but fails to be regular on the
event horizon, whilst the other is regular on the
event horizon but not invariant under simultaneous $t$,
$\phi $ reversal.
We should add that are conclusions are based on a \emph{mode 
by mode} analysis and it is possible, though in our opinion
unlikely, that subtle cancellations could rescue the 
Frolov-Thorne state.

A detailed numerical investigation would be necessary
to elucidate further details of the properties
of these states outside the event horizon.
This paper has laid the foundation for such an investigation 
which we will present in  following papers in this series.

It is possible to draw some conclusions on
the basis of our analysis without resorting to a numerical
investigation.  For example, one can show that any state which is 
isotropic in a tetrad which co-rotates with the event horizon
must become divergent on the velocity of light surface 
\cite{PLApreprint}.
This implies that even if we could construct a state
which is regular on the event horizon and has the
desired thermal properties, then that state may well
turn out not to be regular on the velocity of 
light surface, in agreement with the Kay-Wald theorem
that the state must fail to be regular somewhere.

This paper has shown that whilst quantum field 
theory in Kerr space-time is more complex than in 
Schwarzschild, application of the same physical
principles which have proved to be so valuable
in Schwarzschild also makes the picture
much clearer and more simple in Kerr.

\acknowledgements
E.W. thanks Oriel College, Oxford, for a fellowship supporting
this research, and the Department of Physics, University of
Newcastle, Newcastle-upon-Tyne, for hospitality during the completion
of this work.

\end{document}